\documentclass[conference, a4paper]{IEEEtran}
\IEEEoverridecommandlockouts

	\usepackage[turkish]{babel}
	\usepackage[utf8]{inputenc} 
	\usepackage[T1]{fontenc}

\usepackage{cite}

\ifCLASSINFOpdf
  \usepackage[pdftex]{graphicx}
\else
\fi
\usepackage[cmex10]{amsmath}
\usepackage{multirow}
\usepackage{array}
\usepackage[lofdepth,lotdepth]{subfig}

\AtBeginDocument{%
  \renewcommand\tablename{TABLO}
}

\AtBeginDocument{%
  \renewcommand\abstractname{Abstract}
}

\begin{document}


%
\title{Akıllı Yansıtıcı Yüzey Destekli Çok-Girişli Tek-Çıkışlı Sistemde Kanal Kestirimi\\
Channel Estimation for RIS aided MISO System}

\author{\IEEEauthorblockN{\textit{Rıfat Volkan Şenyuva}}
\IEEEauthorblockA{\textit{Elektrik-Elektronik Mühendisliği Bölümü} \\
\textit{Maltepe Üniversitesi}\\
İstanbul, Türkiye \\
rifatvolkansenyuva@maltepe.edu.tr}}


%

\maketitle

\begin{ozet}
Bu çalışmada akıllı yansıtıcı yüzey (AYY) destekli çok-girişli tek-çıkışlı tek kullanıcılı sistemde kanal kestirimi incelenmektedir. Ayrık Fourier dönüşümü matrisine göre her sembol aralığında etkinleştirilen AYY elemanlarını kullanan en küçük varyanslı yansız (EKVY) kestiricisi ile en küçük ortalama kareler (EKOK) kestiricisinin başarımları karşılaştırılmaktadır. EKOK kestiricisiyle, EKVY kestiricisi üzerine sinyal gürültü oranında 10 dB üzerinde iyileştirme elde edilebileceği sayısal sonuçlarla gösterilmektedir. 
\end{ozet}
\begin{IEEEanahtar}
akıllı yansıtıcı yüzeyler, kanal kestirimi, en küçük varyanslı yansız kestiricisi, en küçük ortalama kareler kestiricisi.
\end{IEEEanahtar}

\begin{abstract}
This paper considers the channel estimation of a single user in a MISO system with an intelligent reflecting surface (IRS). The performances of the minimum variance unbiased (MVU) and minimum mean square error (MMSE) estimators using the discrete Fourier transform activation pattern for the IRS, updated at every symbol interval, are compared. Numerical results show that the MMSE estimator provides over 10 dB SNR improvement compared to the MVU estimator.
\end{abstract}



%
\IEEEpeerreviewmaketitle

\IEEEpubidadjcol

\section{G{\footnotesize İ}r{\footnotesize İ}ş}

5. nesil ve sonrası kablosuz ağlarda artan veri hızları nedeniyle enerji verimliliği ön plana çıkmaktadır. Çok antenli baz istasyonları ve erişim noktalarının yoğun kurulumuyla tesis edilecek bu yeni nesil ağlara 50 milyardan fazla cihazın bağlanması beklenmektedir \cite{energyEfficiency}. Güç tüketimlerinin çok fazla olması ve kullanıcılarına kesintisiz yüksek nitelikli hizmet sağlamakta zorlanmaları bu yeni nesil kablosuz ağ teknolojilerinin gerçeklenmesi önündeki en önemli iki problemdir \cite{mmse}. Bu problemlerin çözümü için radyo yayılım ortamının akıllı yansıtıcı yüzey (AYY) kullanılarak akıllı hale getirilmesi önerilmektedir \cite{SRE}. AYY gelen elektromanyetik alana pasif huzme şekillendirme uygulayabilen ve bütünleşik elektronik devreler aracılığıyla programlanabilen meta yüzeylerdir \cite{SRE}. AYY' de rölelerde olduğu gibi güç yükselteçleri kullanılmadığı için çok daha az güç tüketilmektedir. Bu nedenle enerji verimliliğini artırmayı hedefleyen yeni nesil ağ teknolojilerinde AYY' in kullanılması öngörülmektedir. 

Yalnız bu teknolojinin olgunlaşabilmesi için sinyal işleme alanında çözülmesi gereken problemler vardır. Bu problemlerden birisi kanal kestirimidir. AYY' de artan eleman sayısıyla birlikte kestirilmesi gereken kanal sayısı artmakta, pasif elemanlardan oluşmaları nedeniyle kestirim ancak alıcı tarafında yapılabilmekte ve geleneksel sistemlere göre daha yüksek kestirim hatası ortaya çıkmaktadır \cite{mvu}. Kanal kestirimini iyileştirmek amacıyla AYY elemanlarının nasıl etkinleştirilmesi gerektiği üzerine araştırmalar yapılmaktadır \cite{beamformingWirelessTransfer,mvu,mmse, parafacApproach, chEstIRS}. \cite{beamformingWirelessTransfer} ' te AYY elemanlarının bir kısmının açık, geri kalanlarının ise kapalı bırakılarak en küçük varyans yansız (EKVY) kestiricisi uygulanmaktayken, \cite{mvu}' de ise elemanların Ayrık Fourier Dönüşümü (AFD) matrisine göre etkinleştirilerek uygulanan EKVY kestiricisinin varyansının kılavuz sembol sayısıyla azaltılabileceği gösterilmektedir. \cite{mmse}' de EKVY kestiricisi üzerine en küçük ortalama kareler (EKOK) kestiricisi geliştirilmesi anlatılmaktadır. \cite{mvu,beamformingWirelessTransfer}' te AYY elemanlarının değiştirilme sıklığı sembol aralığı olarak belirlenmekteyken \cite{mmse, parafacApproach, chEstIRS}' deki sinyal modelinde ise kanal uyum süresinde içerisinde kalmak şartıyla belli sayıda sembol gönderimi için AYY elemanları aynı kalmaktadır.

Bu çalışmada AYY tarafından desteklenen çok-girişli tek-çıkışlı sistemde tek kullanıcı için kanal kestirimi problemi ele alınmaktadır. Sinyal modelinde AYY ile baz istasyonu arasındaki kanalların gerekirci saçılmayla, diğer kanalların ise bağımsız Rayleigh sönümlemesine göre modellenebildiği ve AYY elemanlarının kılavuz sembol aralığında değiştirilebildiği kabul edilmektedir. Bu sinyal modelinde önce AFD matrisine göre etkinleştirilen elemanları kullanan EKVY kestiricisi ve kovaryans matrisi türetimleri verilmekte sonrasında EKVY üzerine EKOK kestiricisi ve kovaryans matrisi gösterilmektedir. Sayısal sonuçlarda her iki kestiricinin başarımları karşılaştırılmaktadır.

\section{S{\footnotesize İ}nyal Model{\footnotesize İ}}
İncelenen çok-girişli tek-çıkışlı iletişim sistemi, $N$ elemanlı AYY tarafından desteklenmektedir. AYY elemanları, $\boldsymbol{\phi}_{t}=[\phi_{t,1},...,\phi_{t,N}]^{T}$, yüzeye gelen sinyallerin genliklerini ve fazlarını, $\phi_{t,n}=\alpha_{t,n}\exp(i\theta_{t,n})$ çarpanlarıyla etkilemektedir. Çarpanların genliği $\alpha_{t,n}\in[0,1]$ ve fazları $\theta_{t,n}\in[0,2\pi)$ verilen aralıklar içerisinde her $t$ sembol aralığında kontrol kanalı üzerinden değiştirilebilmektedir \cite{mvu}. Bu sistemde $M$ antenli bir baz istasyonu, tek antenli bir kullanıcıya hizmet vermektedir. İletişim zaman bölüşümlü yarı çift yönlü gerçekleştiğinden kanal uyum süresi içerisinde baz istasyonu-kullanıcı ve kullanıcı-baz istasyonu bağlantı kanalları aynı kalmaktadır. Böylece kullanıcı-baz istasyonu bağlantısı yönünde baz istasyonunda yapılacak kanal kestirimiyle, baz istasyonu-kullanıcı bağlantısı yönünde sembollerin kodlanarak gönderilmesi mümkün olabilmektedir. 

Kanal kestiriminde kullanıcı tarafından her $t$ sembol aralığında bir kılavuz sembol gönderilmekte ve gönderilen bu kılavuz sembol
\begin{equation}
    \mathbf{y}_{t}=(\mathbf{h}_{\text{BS}}+\mathbf{H}_{\text{BS-RIS}}\mathbf{\Phi}_{t}\mathbf{h}_{\text{IRS}})x_{t}+\mathbf{z}_{t}
    \label{eq:uplinkOneSymbol}
\end{equation}
$\mathbf{y}_{t}\in\mathbf{C}^{M}$ olarak baz istasyonuna iletilmektedir \cite{mvu,mmse}. Denklem \eqref{eq:uplinkOneSymbol}' de $\mathbf{h}_{\text{BS}}\in\mathbf{C}^{M}$ baz istasyonu-kullanıcı direk bağlantı kanalını, $\mathbf{H}_{\text{BS-IRS}}\in\mathbf{C}^{M\times N}$ matrisi baz istasyonu ve AYY arasındaki kanalları, $\mathbf{\Phi}_{t}=\text{diag}\lbrace \boldsymbol{\phi}_{t} \rbrace \in\mathbf{C}^{N\times N}$ $t$ sembol aralığındaki AYY çarpanları matrisini, $\mathbf{h}_{\text{IRS}}\in\mathbf{C}^{N}$ kullanıcı ile AYY arasındaki kanalları, $x_{t}\in \mathbf{C}$ kılavuz sembolü, $\mathbf{z}_{t}\sim\mathcal{CN}(\mathbf{0},N_{0}\mathbf{I}_{M})$ ise tek taraflı spektral güç yoğunluğu $N_{0}$ olan toplamsal karmaşık Gauss gürültüsünü göstermektedir. Kullanıcıyla baz istasyonu ve AYY ile arasındaki kanallar için
\begin{IEEEeqnarray}{rCl}
    \mathbf{h}_{\text{BS}}=[h_{\text{BS}_{1}} \cdots h_{\text{BS}_{M}}]^{T}&=&\sqrt{\beta_{\text{BS}}}\mathbf{g}_{\text{BS}} \label{eq:directBSCh} \\
    \mathbf{h}_{\text{IRS}}=[h_{\text{IRS}_{1}} \cdots h_{\text{IRS}_{N}}]^{T}&=&\sqrt{\beta_{\text{IRS}}}\mathbf{g}_{\text{IRS}}
    \label{eq:IRSCh}
\end{IEEEeqnarray}
bağımsız Rayleigh sönümlemesi modeli kullanılmaktadır \cite{mmse,mvu}. Denklem \eqref{eq:directBSCh} ve \eqref{eq:IRSCh}' te yol kayıpları $\beta_{\text{BS}}$ ve $\beta_{\text{IRS}}$ ile, çok yollu sönümleme ise, $\mathbf{g}_{\text{BS}}\sim\mathcal{CN}(\mathbf{0},\mathbf{I}_{M})$ ve $\mathbf{g}_{\text{IRS}}\sim\mathcal{CN}(\mathbf{0},\mathbf{I}_{N})$ bağımsız karmaşık Gauss değişkenleriyle temsil edilmektedir. AYY' nin baz istasyonuna yakın ve görüş hattında olduğu kabul edilerek $\mathbf{H}_{\text{BS-IRS}}$ için yüksek kerteli gerekirci saçılma modeli kullanılmaktadır \cite{aymptoticSINR}. Yüksek kerteli $\mathbf{H}_{\text{BS-IRS}}$ kanal matrisi
\begin{IEEEeqnarray}{rCl}
    &&\left\lbrack \mathbf{H}_{\text{BS-IRS}}\right\rbrack_{m,n}=\sqrt{\beta_{\text{BS-IRS}}}\exp\bigg\lbrack i\frac{2\pi}{\lambda}(m-1)d_{\text{BS}}\sin(\vartheta_{\text{LoSD}_{n}})\nonumber \\
    &&\sin(\varphi_{\text{LoSD}_{n}})+(n-1)d_{\text{IRS}}\sin(\vartheta_{\text{LoSA}_{m}})\sin(\varphi_{\text{LoSA}_{m}})\bigg\rbrack
    \label{eq:BS-IRS_Ch}
\end{IEEEeqnarray}
şeklinde $m=1,...,M$ ve $n=1,...,N$ için oluşturulmaktadır. Denklem \eqref{eq:BS-IRS_Ch}' te yol kayıpları $\beta_{\text{BS-IRS}}$, taşıyıcı dalga boyu $\lambda$, baz istasyonu antenleri ve akıllı yansıtıcı yüzey elemanları arasındaki mesafeler sırasıyla $d_{\text{BS}}$ ve $d_{\text{IRS}}$, baz istasyonundan ayrılış yükseklik ve azimut açıları sırasıyla $\vartheta_{\text{LoSD}_{n}}$ ve $\varphi_{\text{LoSD}_{n}}$ ve son olarak AYY' ye geliş yükseklik ve azimut açıları sırasıyla $\vartheta_{\text{LoSA}_{m}}$ ve $\varphi_{\text{LoSA}_{m}}$ ile temsil etmektedir.

$\mathbf{\Phi}_{t}\mathbf{h}_{\text{IRS}}$ çarpımı yerine $\text{diag}\left\lbrace\mathbf{h}_{\text{IRS}}\right\rbrace\boldsymbol{\phi}_{t}$ kullanılarak denklem \eqref{eq:uplinkOneSymbol} tekrar yazılırsa 
\begin{IEEEeqnarray}{rCl}
    \mathbf{y}_{t}&=&(\mathbf{h}_{\text{BS}}+\mathbf{H}_{\text{BS-RIS}}\text{diag}\left\lbrace\mathbf{h}_{\text{IRS}}\right\rbrace\boldsymbol{\phi}_{t})x_{t}+\mathbf{z}_{t} \label{eq:uplinkOneSymbolSecondForm} \\
    &=&[\mathbf{h}_{\text{BS}} \,\, \mathbf{H}_{\text{cascade}}]\left\lbrack \begin{array}{c} 1 \\ \boldsymbol{\phi}_{t} \end{array} \right\rbrack x_{t}+\mathbf{z}_{t}
    \label{eq:uplinkOneSymbolThirdForm}
\end{IEEEeqnarray}
şeklinde \eqref{eq:uplinkOneSymbolSecondForm} ve \eqref{eq:uplinkOneSymbolThirdForm} denklemleri elde edilebilmektedir. Denklem \eqref{eq:uplinkOneSymbolThirdForm}' da verilen kaskat kanalın sütunları
\begin{IEEEeqnarray}{rCl}
    \mathbf{H}_{\text{cascade}}&=&\mathbf{H}_{\text{BS-RIS}}\text{diag}\left\lbrace\mathbf{h}_{\text{IRS}}\right\rbrace \\
    &=&\left\lbrack \begin{array}{ccc}
         h_{\text{IRS}_{1}}\mathbf{h}_{\text{BS-RIS}_{1}} & \cdots & h_{\text{IRS}_{N}}\mathbf{h}_{\text{BS-RIS}_{N}}
    \end{array}\right\rbrack
    \label{eq:cascadeCh}
\end{IEEEeqnarray}
olarak yazılabilmektedir. Kılavuz semboller $|x_{t}|^{2}=1$ sağlayacak şekilde seçilir ve denklem \eqref{eq:uplinkOneSymbolThirdForm}' nın her iki tarafı kılavuz sembolün eşleniği, $x_{t}^{*}$, ile çarpılırsa
\begin{equation}
    \widetilde{\mathbf{y}}_{t}=[\mathbf{h}_{\text{BS}} \,\, \mathbf{H}_{\text{cascade}}]\left\lbrack \begin{array}{c} 1 \\ \boldsymbol{\phi}_{t} \end{array} \right\rbrack +\widetilde{\mathbf{z}}_{t}
    \label{eq:uplinkOneSymbolDespreading}
\end{equation}
denklemi elde edilmektedir. Denklem \eqref{eq:uplinkOneSymbolDespreading}' da toplam sinyal $\widetilde{\mathbf{y}}_{t}=x_{t}^{*}\mathbf{y}_{t}$, gürültü bileşeni ise $\widetilde{\mathbf{z}}_{t}=x_{t}^{*}\mathbf{z}_{t}$ olarak verilmektedir. Yeni gürültü bileşeninin dağılımı, $\widetilde{\mathbf{z}}_{t}\sim \mathcal{CN}(\mathbf{0},N_{0}\mathbf{I}_{M})$, eskisiyle \eqref{eq:uplinkOneSymbol} aynıdır. Kanal uyum süresi içerisinde toplam $\tau_{p}$ adet kılavuz sembol için baz istasyonundaki sinyallerin toplamı
\begin{IEEEeqnarray}{rCl}
    \widetilde{\mathbf{Y}}&=&[\mathbf{h}_{\text{BS}} \,\, \mathbf{H}_{\text{cascade}}] \left\lbrack \begin{array}{ccc} 1 & \cdots & 1 \\ \boldsymbol{\phi}_{1} & \cdots & \boldsymbol{\phi}_{\tau_{p}} \end{array} \right\rbrack+\widetilde{\mathbf{Z}}
    \label{eq:uplinkAllSymbols} \\
    &=&\mathbf{H}_{\text{comp}}\mathbf{V}+\widetilde{\mathbf{Z}}
    \label{eq:uplinkAllSymCompCh}
\end{IEEEeqnarray}
şeklinde yazılabilmektedir. Denklem \eqref{eq:uplinkAllSymbols}' da toplam sinyal matrisi $\widetilde{\mathbf{Y}}=[\widetilde{\mathbf{y}}_{1} \cdots \widetilde{\mathbf{y}}_{\tau_{p}}]\in\mathbf{C}^{M\times \tau_{p}}$, toplam gürültü matrisi ise $\widetilde{\mathbf{Z}}=[\widetilde{\mathbf{z}}_{1}\cdots\widetilde{\mathbf{z}}_{\tau_{p}}]\in\mathbf{C}^{M\times\tau_{p}}$ olarak verilmektedir. Denklem \eqref{eq:uplinkAllSymCompCh}' de bileşik kanal matrisi $\mathbf{H}_{\text{comp}}\in\mathbf{C}^{M\times (N+1)}$ ile temsil edilmekteyken yeni AYY elemanları matrisi ise $\mathbf{V}\in\mathbf{C}^{(N+1)\times \tau_{p}}$ ile gösterilmektedir. Baz istasyonu-kullanıcı bağlantısı yönünde sembollerin kodlanarak iletilebilmesi için bileşik kanal matrisinin kestirilmesi gerekmektedir.




\section{En Küçük Varyanslı Yansız Kest{\footnotesize İ}r{\footnotesize İ}c{\footnotesize İ}s{\footnotesize İ}}

Denklem \eqref{eq:uplinkAllSymCompCh}' in her iki tarafı vektör haline getirilirse
\begin{IEEEeqnarray}{rCl}
    \text{vec}\lbrace \widetilde{\mathbf{Y}} \rbrace &=& \text{vec}\lbrace \mathbf{H}_{\text{comp}}\mathbf{V} \rbrace+\text{vec}\lbrace \widetilde{\mathbf{Z}}\rbrace \label{eq:vectorForm} \\
    \widetilde{\mathbf{y}}&=& (\widetilde{\mathbf{V}} \otimes \mathbf{I}_{M}) \text{vec}\lbrace \mathbf{H}_{\text{comp}}\rbrace+\widetilde{\mathbf{z}} \label{eq:kronProd} \\
    &=&\mathbf{A}\mathbf{h}+\widetilde{\mathbf{z}} \label{eq:linearModel}
\end{IEEEeqnarray}
şeklinde baz istasyonundaki toplam sinyal denklem sistemine dönüştürülebilmektedir. Denklem \eqref{eq:vectorForm}' dan denklem \eqref{eq:kronProd}' e geçişte Kronecker çarpım, $\otimes$, özelliği olan $\text{vec}\lbrace \mathbf{I}_{M}\mathbf{H}_{\text{comp}}\mathbf{V}\rbrace=(\widetilde{\mathbf{V}} \otimes \mathbf{I}_{M}) \text{vec}\lbrace \mathbf{H}_{\text{comp}}\rbrace$ kullanılmaktadır ve $\mathbf{V}$ matrisinin devriği $\widetilde{\mathbf{V}}=\mathbf{V}^{T}$ olarak temsil edilmektedir. Denklem \eqref{eq:linearModel}' te $\mathbf{A}$ matrisi $\mathbf{A}=\widetilde{\mathbf{V}} \otimes \mathbf{I}_{M}$ olarak verilmekteyken bileşik kanal vektörü
\begin{IEEEeqnarray}{rCl}
    \mathbf{h}&=&\left\lbrack \begin{array}{cccc}
                    \mathbf{h}_{\text{BS}}^{T} & \mathbf{h}_{\text{cascade}_{1}}^{T} & \cdots & \mathbf{h}_{\text{cascade}_{N}}^{T}
                  \end{array} \right\rbrack^{T} \label{eq:compCh}
\end{IEEEeqnarray}
şeklinde baz istasyonu-kullanıcı arasındaki direk kanal vektörüyle kaskat kanal matrisi sütunlarının bitiştirilmesiyle oluşturulmaktadır. Bileşik kanal vektörü \eqref{eq:compCh}, $\mathbf{h}\sim\mathcal{CN}(\mathbf{0},\mathbf{C}_{\mathbf{H}})$, kovaryans matrisi
\begin{IEEEeqnarray}{rCl}
    \mathbf{C}_{\mathbf{H}}&=&\text{blkdiag}\lbrace \beta_{\text{BS}}\mathbf{I}_{M},\beta_{\text{IRS}}\mathbf{h}_{\text{BS-RIS}_{1}}\mathbf{h}_{\text{BS-RIS}_{1}}^{H},... \nonumber \\
    &,&\beta_{\text{IRS}}\mathbf{h}_{\text{BS-RIS}_{N}}\mathbf{h}_{\text{BS-RIS}_{N}}^{H} \rbrace
\end{IEEEeqnarray}
$M(N+1)\times M(N+1)$ boyutlu blok köşegen matris ile verilmekteyken gürültü bileşeni \eqref{eq:linearModel} dağılımı $\widetilde{\mathbf{z}}\sim \mathcal{CN}(\mathbf{0},N_{0}\mathbf{I}_{M\tau_{p}})$ olmaktadır. Kılavuz sembol sayısı, AYY eleman sayısından bir fazla seçilirse, $\tau_{p}\ge N+1$, denklem \eqref{eq:linearModel}' te $\mathbf{h}$ vektörü için EKVY kesitiricisi
\begin{IEEEeqnarray}{rCl}
    \hat{\mathbf{h}}_{\text{mvu}}&=&\text{argmin}\,E\lbrace \|\mathbf{A}\mathbf{h}-\widetilde{\mathbf{y}}\|_{2}^{2} \rbrace \\
    &=&(\mathbf{A}^{H}\mathbf{A})^{-1}\mathbf{A}^{H}\widetilde{\mathbf{y}} \\
    &=&[(\widetilde{\mathbf{V}} \otimes \mathbf{I}_{M})^{H}(\widetilde{\mathbf{V}} \otimes \mathbf{I}_{M})]^{-1}(\widetilde{\mathbf{V}} \otimes \mathbf{I}_{M})^{H}\widetilde{\mathbf{y}} \\
    &=&(\widetilde{\mathbf{V}}^{H}\widetilde{\mathbf{V}}\otimes\mathbf{I}_{M})^{-1}(\widetilde{\mathbf{V}}^{H} \otimes \mathbf{I}_{M})\widetilde{\mathbf{y}} \\
    &=&\lbrack(\widetilde{\mathbf{V}}^{H}\widetilde{\mathbf{V}})^{-1}\otimes\mathbf{I}_{M}\rbrack(\widetilde{\mathbf{V}}^{H} \otimes \mathbf{I}_{M})\widetilde{\mathbf{y}}
    \label{eq:mvuEstimator}
\end{IEEEeqnarray}
en küçük kareler (EKK) kestiricisi olmaktadır. Denklem \eqref{eq:mvuEstimator}' de $\widetilde{\mathbf{V}}^{H}$ matrisi $\widetilde{\mathbf{V}}$ matrisinin Hermit eşleniğini, $(\widetilde{\mathbf{V}}^{H}\widetilde{\mathbf{V}})^{-1}$ ise $\widetilde{\mathbf{V}}^{H}\widetilde{\mathbf{V}}$ matrisinin tersini göstermektedir. Denklem \eqref{eq:mvuEstimator} görüleceği üzere EKK kestiricisinin başarımını $\widetilde{\mathbf{V}}$ matrisi belirlemektedir. AYY elemanlarının bir kısmının etkinleştirilip kalanların ise kapalı, $\phi_{t,n}\in\lbrace 0,1 \rbrace$, bırakıldığı zaman \cite{beamformingWirelessTransfer}
\begin{equation}
    \widetilde{\mathbf{V}}_{\text{onoff}}=\left\lbrack \begin{array}{cc}
        1 & \mathbf{0}_{N}^{T}  \\
        \mathbf{1}_{N} & \mathbf{I}_{N} 
    \end{array} \right\rbrack
    \label{eq:risOnOffPattern}
\end{equation}
ortaya çıkan $\widetilde{\mathbf{V}}_{\text{onoff}}$ matrisini \eqref{eq:risOnOffPattern} kullanan kestirici
\begin{equation}
    \hat{\mathbf{h}}_{\text{mvu-onoff}}=\left( \widetilde{\mathbf{V}}^{-1}_{\text{onoff}} \otimes \mathbf{I}_{M} \right)\widetilde{\mathbf{y}}
    \label{eq:mvuOnOff}
\end{equation}
için direk kanal kestirim hatasının kaskat kanal kestirim hatasını artırdığı tespit edilmektedir \cite{mvu}. 

EKK kestiricisinin kovaryans matrisi, $\mathbf{C}_{\hat{\mathbf{H}}_{\text{mvu}}}$, köşegen olduğunda Fisher matrisinin, $\mathcal{I}(\mathbf{h})$, tersine eşitlenmekte
\begin{IEEEeqnarray}{rCl}
    \mathbf{C}_{\hat{\mathbf{H}}_{\text{mvu}}}=\mathcal{I}^{-1}(\mathbf{h})&=&N_{0}(\mathbf{A}^{H}\mathbf{A})^{-1} \label{eq:cramerRao} \\
    &=&N_{0}\lbrack(\widetilde{\mathbf{V}}^{H}\widetilde{\mathbf{V}})^{-1}\otimes\mathbf{I}_{M}\rbrack \label{eq:lsCovariance}
\end{IEEEeqnarray}
 ve $\mathbf{h}$ bileşik kanal vektörü kestirimi için Cramer Rao hata alt sınırına ulaşılmaktadır. EKK kestiricisi kovaryans matrisinin \eqref{eq:lsCovariance} izi kestiricinin ortalama karesel hatasını (OKH) belirlediğinden $\widetilde{\mathbf{V}}$ matrisi seçiminin, kovaryans matrisi izini en küçük yapacak şekilde yapılması gerekmektedir. Kovaryans matrisinin \eqref{eq:lsCovariance} köşegen olması için $\widetilde{\mathbf{V}}^{H}\widetilde{\mathbf{V}}$ matrisinin de köşegen olması gerekmektedir \cite{mvu}. $\widetilde{\mathbf{V}}^{H}\widetilde{\mathbf{V}}$ matrisinin izi
 \begin{equation}
     \text{tr}\lbrace \widetilde{\mathbf{V}}^{H}\widetilde{\mathbf{V}}\rbrace\le \tau_{p}(N+1)
     \label{eq:trace}
 \end{equation}
 üstten sınırlı olduğundan alabileceği en küçük değer $\tau_{p}(N+1)$ olmaktadır. $\widetilde{\mathbf{V}}$ matrisi, $\tau_{p}\times \tau_{p}$ boyutlu AFD matrisinin ilk $N+1$ sütunundan seçilirse, $\mathbf{F}\in\mathbf{C}^{\tau_{p}\times (N+1)}$, $\widetilde{\mathbf{V}}^{H}\widetilde{\mathbf{V}}$ matrisinin izi \eqref{eq:trace}, $\tau_{p}(N+1)$ değerine eşitlenmektedir. Bu seçim sonucunda kovaryans matrisi
 \begin{IEEEeqnarray}{rCl}
    \mathbf{C}_{\hat{\mathbf{H}}_{\text{mvu}}}&=&N_{0}\lbrack(\mathbf{F}^{H}\mathbf{F})^{-1}\otimes\mathbf{I}_{M}\rbrack \\
    &=&N_{0}\lbrack (\tau_{p}\mathbf{I}_{N+1})^{-1}\otimes\mathbf{I}_{M} \rbrack \\
    &=&\frac{N_{0}}{\tau_{p}}\mathbf{I}_{M(N+1)} \label{eq:lsDFTCovar}
 \end{IEEEeqnarray}
olarak hesaplanabilmektedir \cite{mvu}. Denklem \eqref{eq:lsDFTCovar}' den görüleceği üzere AYY çarpanları AFD matrisinden seçilirse kestirim varyansı kılavuz sembol sayısı ile ters orantılı çıkmaktadır. Böylece kılavuz sembol sayısı artırılarak kestirim varyansını düşürmek mümkün olabilmektedir. Bu seçim sonucu elde edilecek EKK kestiricisi 
\begin{equation}
    \hat{\mathbf{h}}_{\text{mvu-dft}}=\frac{1}{\tau_{p}}(\mathbf{F}^{H} \otimes \mathbf{I}_{M})\widetilde{\mathbf{y}}
    \label{eq:lsDFT}
\end{equation}
olarak ifade edilebilmektedir. Denklem \eqref{eq:lsDFT}' teki EKK kestiricisinin normalize OKH baz istasyonu-kullanıcı arasındaki direk kanal vektörü için
\begin{equation}
    \text{NMSE}_{\text{mvu-bs}}=\frac{\text{tr}\lbrace (N_{0}/\tau_{p})\mathbf{I}_{M}  \rbrace}{\text{tr}\lbrace \beta_{\text{BS}} \mathbf{I}_{M}\rbrace}=\frac{N_{0}}{\tau_{p}\beta_{\text{BS}}}
    \label{eq:nmseLSBS}
\end{equation}
olarak bulunabilmektedir. Kaskat kanal vektörleri için EKK kestiricisi normalize OKH ise
\begin{equation}
    \text{NMSE}_{\text{mvu-cas}_{n}}=\frac{\text{tr}\lbrace (N_{0}/\tau_{p})\mathbf{I}_{M} \rbrace}{\text{tr}\lbrace \beta_{\text{IRS}}\mathbf{h}_{\text{BS-RIS}_{n}}\mathbf{h}_{\text{BS-RIS}_{n}}^{H} \rbrace}=\frac{N_{0}}{\tau_{p}\beta_{\text{IRS}}\beta_{\text{BS-IRS}}}
    \label{eq:nmseLSCascade}
\end{equation}
şeklinde hesaplanmaktadır.
 
\section{En Küçük Ortalama Kareler Kest{\footnotesize İ}r{\footnotesize İ}c{\footnotesize İ}s{\footnotesize İ}}
EKOK kestiricisi için önce EKVY kestiricisi \eqref{eq:lsDFT} 
\begin{IEEEeqnarray}{rCl}
    \mathbf{r}&=&\frac{1}{\tau_{p}}(\mathbf{F}^{H} \otimes \mathbf{I}_{M})\widetilde{\mathbf{y}} \\
              &=&\frac{1}{\tau_{p}}(\mathbf{F}^{H} \otimes \mathbf{I}_{M})\lbrack(\mathbf{F} \otimes \mathbf{I}_{M}) \mathbf{h}+\widetilde{\mathbf{z}}\rbrack \\
              &=&\frac{1}{\tau_{p}}(\mathbf{F}^{H}\mathbf{F} \otimes \mathbf{I}_{M})\mathbf{h}+\frac{1}{\tau_{p}}(\mathbf{F}^{H} \otimes \mathbf{I}_{M})\widetilde{\mathbf{z}} \\
              &=& \mathbf{h}+\mathbf{w} 
              \label{eq:lsDFTApplied}
\end{IEEEeqnarray}
şeklinde uygulanmaktadır \cite{mmse}. Denklem \eqref{eq:lsDFTApplied}' te ortaya çıkan gürültü vektörünün dağılımı $\mathbf{w}\sim\mathcal{CN}(\mathbf{0},\frac{N_{0}}{\tau_{p}}\mathbf{I}_{M(N+1)})$. 

Denklem \eqref{eq:lsDFTApplied} için EKOK kestiricisi
\begin{equation}
    \hat{\mathbf{h}}_{\text{mmse}}=\mathbf{C}_{\mathbf{H}\mathbf{R}}\mathbf{C}^{-1}_{\mathbf{R}}\mathbf{r}
    \label{eq:mmseEstimatorI}
\end{equation}
olarak verilmektedir \cite{gallager}. $\mathbf{H}\mathbf{R}$ bileşik değişkeni için kovaryans matrisi, $\mathbf{C}_{\mathbf{H}\mathbf{R}}=\mathbf{C}_{\mathbf{H}}$ iken $\mathbf{R}$ değişkeni kovaryans matrisi ise $\mathbf{C}_{\mathbf{R}}=\mathbf{C}_{\mathbf{H}}+\frac{N_{0}}{\tau_{p}}\mathbf{I}_{M(N+1)}$ olmaktadır. Kovaryans matrisleri denklem \eqref{eq:mmseEstimatorI}' de yerine yazılırsa EKOK kestiricisi
\begin{equation}
    \hat{\mathbf{h}}_{\text{mmse}}=\mathbf{C}_{\mathbf{H}}\left\lbrack\mathbf{C}_{\mathbf{H}}+(N_{0}/\tau_{p})\mathbf{I}_{M(N+1)}\right\rbrack^{-1}\mathbf{r}
    \label{eq:mmseEstimatorII}
\end{equation}
formunu almaktadır. Baz istasyonu-kullanıcı arasındaki direk kanal vektörü, kovaryans matrisi $\beta_{\text{BS}}$ ile ölçeklenmiş birim matris olduğundan sadece $\mathbf{h}_{\text{BS}}$ kanal vektörü için EKOK kestiricisi
\begin{equation}
    \hat{\mathbf{h}}_{\text{mmse-bs}}=\left(\frac{\beta_{\text{BS}}}{\beta_{\text{BS}}+N_{0}/\tau_{p}}\right)\mathbf{r}_{1:M}
    \label{eq:mmseBS}
\end{equation}
şeklinde basitleşebilmektedir. Denklem \eqref{eq:mmseBS}' te $\mathbf{r}_{1:M}$ ile $\mathbf{r}$ vektörünün ilk $M$ elemanından oluşan vektör temsil edilmektedir. Direk kanal vektörü EKOK kestiricisinin \eqref{eq:mmseBS} kovaryans matrisi ise
\begin{equation}
    \mathbf{C}_{\hat{\mathbf{h}}_{\text{mmse-bs}}}=\frac{\beta_{\text{BS}}N_{0}/\tau_{p}}{\beta_{\text{BS}}+N_{0}/\tau_{p}}\mathbf{I}_{M}
\end{equation}
şeklinde elde edilebilmektedir. EKOK direk kanal vektörü kestiricisi \eqref{eq:mmseBS} için normalize OKH
\begin{equation}
    \text{NMSE}_{\text{mmse-bs}}=\frac{\text{tr}\lbrace \mathbf{C}_{\hat{\mathbf{h}}_{\text{mmse-bs}}} \rbrace}{\text{tr}\lbrace \beta_{\text{BS}} \mathbf{I}_{M}\rbrace}=\frac{N_{0}}{\tau_{p}\beta_{\text{BS}}+N_{0}}
    \label{eq:nmseMMSEBS}
\end{equation}
olarak verilebilmektedir. Kaskat kanalların normalize OKH ise
\begin{equation}
    \text{NMSE}_{\text{mmse-cas}}=\frac{N_{0}}{\tau_{p}M\beta_{\text{IRS}}\beta_{\text{BS-IRS}}+N_{0}}
    \label{eq:nmseMMSECascade}
\end{equation}
olmaktadır \cite{mmse}.

\section{Sayısal Sonuçlar}
Anlatılan kestirim yöntemlerinin başarımları, normalize OKH' nin sinyal gürültü oranına göre değişimi baz istasyonu-kullanıcı arasındaki direk kanal ve kaskat kanallar için ayrı ayrı çizdirilerek karşılaştırılmaktadır. $10^{5}$ adet Monte Carlo döngüsü sonucunda hesaplanan normalize OKH' lerin örnek ortalaması alınmaktadır. Sinyal gürültü oranının kanalların sönümlemesinden etkilenmemesi için yol kayıp katsayıları 
\begin{equation}
    \beta_{\text{BS}}+N\beta_{\text{IRS}}\beta_{\text{BS-IRS}}=1
\end{equation}
sağlayacak şekilde tespit edilmektedir. $\mathbf{H}_{\text{BS-IRS}}$ kanal matrisi üretiminde antenler ve elemanlar arasındaki mesafe $d_{\text{BS}}=d_{\text{IRS}}=\lambda/2$ olarak belirlenirken yükseklik ve azimut açıları ise sırasıyla $(0,\pi)$ ve $(0,2\pi)$ aralığında tekdüze dağılımdan seçilmektedir. Kılavuz semboller genlikleri birim, $|x_{t}|^{2}=1$, yapılarak sinyal gürültü oranının gürültünün tek taraflı spektral güç yoğunluğunun tersine, $\text{SNR}=1/N_{0}$, eşit olması sağlanmaktadır. Baz istasyonu anten sayısı $M=10$, AYY eleman sayısı ise $N=50$' dir. Kanal kestirimi için toplam $\tau_{p}=N+1=51$ adet kılavuz sembol gönderilmektedir. 

\begin{figure}[htbp]
 	\centering
 	\shorthandoff{=}
 	\includegraphics[scale=0.6]{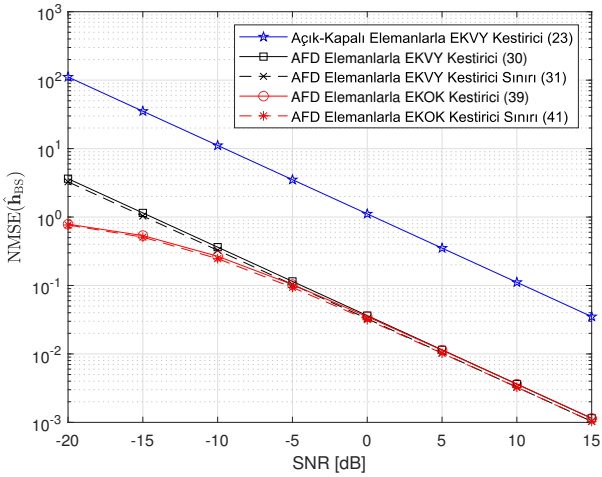}
 	\shorthandon{=}
 	\caption{$M=10$, $N=50$ ve $\tau_{p}=51$ için $\text{NMSE}(\hat{\mathbf{h}}_{\text{BS}})$.}
 	\label{fig:nmseDirect}
\end{figure}

Şekil \ref{fig:nmseDirect}' de sadece baz istasyonu-kullanıcı arasındaki direk kanal için elde edilen normalize OKH' ler gösterilmektedir. AYY elemanlarının bir kısmının açılıp bir kısmının kapatılarak elde edilen $\widetilde{\mathbf{V}}$ matrisini kullanan EKVY kestiricisi, AFD' ye göre elde edilen $\widetilde{\mathbf{V}}$ matrisini kullanan EKVY kestiricisine göre aynı başarım için 15 dB daha fazla sinyal gücü gerektirmektedir. EKVY kestiricisinin üzerine uygulanan EKOK kestircisi ancak -5 dB' den daha küçük sinyal gürültü oranları için iyileştirme sağlayabilmektedir ve bu iyileştirme $\text{SNR=-20}$ dB için 5 dB' ye kadar çıkmaktadır. Şekil \ref{fig:nmseDirect}' den görüleceği üzere EKVY ve EKOK kestirici başarımları sırasıyla denklem \eqref{eq:nmseLSBS} ve denklem \eqref{eq:nmseMMSEBS}' da verilen sınırlarına çok yakın çıkmaktadır.

\begin{figure}[htbp]
 	\centering
 	\shorthandoff{=}
 	\includegraphics[scale=0.6]{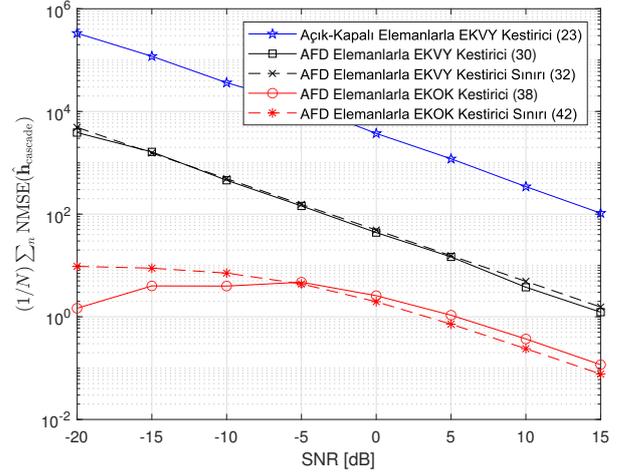}
 	\shorthandon{=}
 	\caption{$M=10$, $N=50$ ve $\tau_{p}=N+1=51$ için $(1/N)\sum_{n}\text{NMSE}(\hat{\mathbf{h}}_{\text{cascade}_{n}})$.}
 	\label{fig:nmseCascade}
\end{figure}

Baz istasyonu-kullanıcı arasındaki kaskat kanal vektörlerinin kestiriminde hesaplanan normalize OKH' lerin ortalaması, $(1/N)\sum_{n}\text{NMSE}(\hat{\mathbf{h}}_{\text{cascade}_{n}})$, Şekil \ref{fig:nmseCascade}' de sunulmaktadır. Direk kanal kestirim hatasına (Şekil \ref{fig:nmseDirect}) kıyasla kaskat kanal kestirim hatasının genel olarak tüm kestiriciler için yükseldiği tespit edilmektedir. EKVY kestiricileri arasındaki sinyal gürültü oranı farkının direk kanal kestirimindeki duruma göre 5 dB artarak 20 dB' ye çıktığı görülmektedir. AFD elemanları kullanan EKOK kestiricisiyle yine AFAD elemanları kullanan EKVY kestiricisine göre -5 dB' den büyük sinyal gürültü oranları için 10 dB üzerinde bir iyileştirme sağlanabileceği mümkün gözükmektedir (Şekil \ref{fig:nmseCascade}).

\section{Sonuç}

Bu çalışmada AYY tarafından desteklenen çok-girişli tek-çıkışlı sistemde bir kullanıcı için EKVY ve EKOK kanal kestirim yöntemlerinin başarımları karşılaştırılmaktadır. Baz istasyonu-AYY arasındaki kanalların gerekirci saçılmayla diğerlerinin ise bağımsız Rayleigh sönümlemesine göre tanımlandığı sinyal modelinde AFD matrisine göre etkinleştirilen AYY için EKVY ve EKOK kanal kestiricilerinin ifadeleri ve bu kestiricilerin kovaryans matrisleri çıkarılmaktadır. Sayısal sonuçlardan görüleceği üzere EKOK kestiricisi kaskat kanalların kestiriminde sinyal gürültü oranında 10 dB üzerinde iyileştirme sağlarken direk kanal kestiriminde ise -5 dB' den büyük sinyal gürültü oranları için EKOK kestiricisi başarımı EKVY kestiriciyle aynıdır.



%

\bibliographystyle{IEEEtran}
\bibliography{IEEEabrv,siu23}

\end{document}